\documentclass[aps,10pt,superscriptaddress,nofootinbib,nobibnotes,amsmath]{revtex4}

\pdfoutput=1

\usepackage{graphicx}

\newcommand{\h}{\overline{\mathrm{H}}}
\newcommand{\anti}[1]{\overline{\mathrm{#1}}}
\newcommand{\pos}{\mathrm{e}^{+}}
\newcommand{\ele}{\mathrm{e}^{-}}
\newcommand{\pic}{\pi^{\pm}}
\newcommand{\pio}{\pi^0}

\begin{document}
\title{Design of a Radial TPC for Antihydrogen Gravity Measurement with ALPHA-g}
\author{A. \surname{Capra}}
\email[Email address: ]{acapra@triumf.ca}
\affiliation{TRIUMF, 4004 Wesbrook Mall, Vancouver BC, Canada V6T 2A3}
\author{P.-A. \surname{Amaudruz}}
\author{D. \surname{Bishop}}
\author{M. C. \surname{Fujiwara}}
\author{S. \surname{Freeman}}
\author{D. \surname{Gill}}
\author{\mbox{M. \surname{Grant}}}
\author{\mbox{R. \surname{Henderson}}}
\author{L. \surname{Kurchaninov}}
\author{P. \surname{Lu}}
\affiliation{TRIUMF, 4004 Wesbrook Mall, Vancouver BC, Canada V6T 2A3}
\author{S. \surname{Menary}}
\affiliation{Department of Physics and Astronomy, York University, Toronto ON, Canada M3J 1P3}
\author{K. \surname{Olchanski}}
\author{F. \surname{Retiere}} 
\affiliation{TRIUMF, 4004 Wesbrook Mall, Vancouver BC, Canada V6T 2A3}


\begin{abstract}
 The gravitational interaction of antimatter and matter has never been directly probed. ALPHA-g is a novel experiment that aims to perform the first measurement of the antihydrogen gravitational mass. A fundamental requirement for this new apparatus is a position sensitive particle detector around the antihydrogen trap which provides information about antihydrogen annihilation location. The proposed detector is a radial Time Projection Chamber, or \textit{rTPC}, whose concept is being developed at TRIUMF. A simulation of the detector and the development of the reconstruction software, used to determine the antihydrogen annihilation point, is presented alongside with the expected performance of the rTPC.
\end{abstract}

\maketitle

\section{Introduction}
\textit{General Relativity} (GR) and the \textit{Standard Model} are incompatible and future experiments will inevitably lead to the modification of one of them, if not both. Testing these theories using antihydrogen ($\h$) is the kind of crucial experiment that is needed. For example, GR does not exhaust the possibilities of metric theories of gravity \cite{wi14cbgre}, whereas the notion of curved spacetime, that follows from the Einstein's Equivalence Principle (EEP), is a very elegant and general one. Therefore, testing the EEP is not an academic exercise but rather it sets the foundation of the modern conception of gravity. Gravitational tests on antimatter are compelling in order to prove the validity of the EEP in the realm of atomic antimatter \cite{ni91aaagaa,ch94ap}.

\begin{figure}[h!]
 \centering
 \includegraphics[scale=0.1]{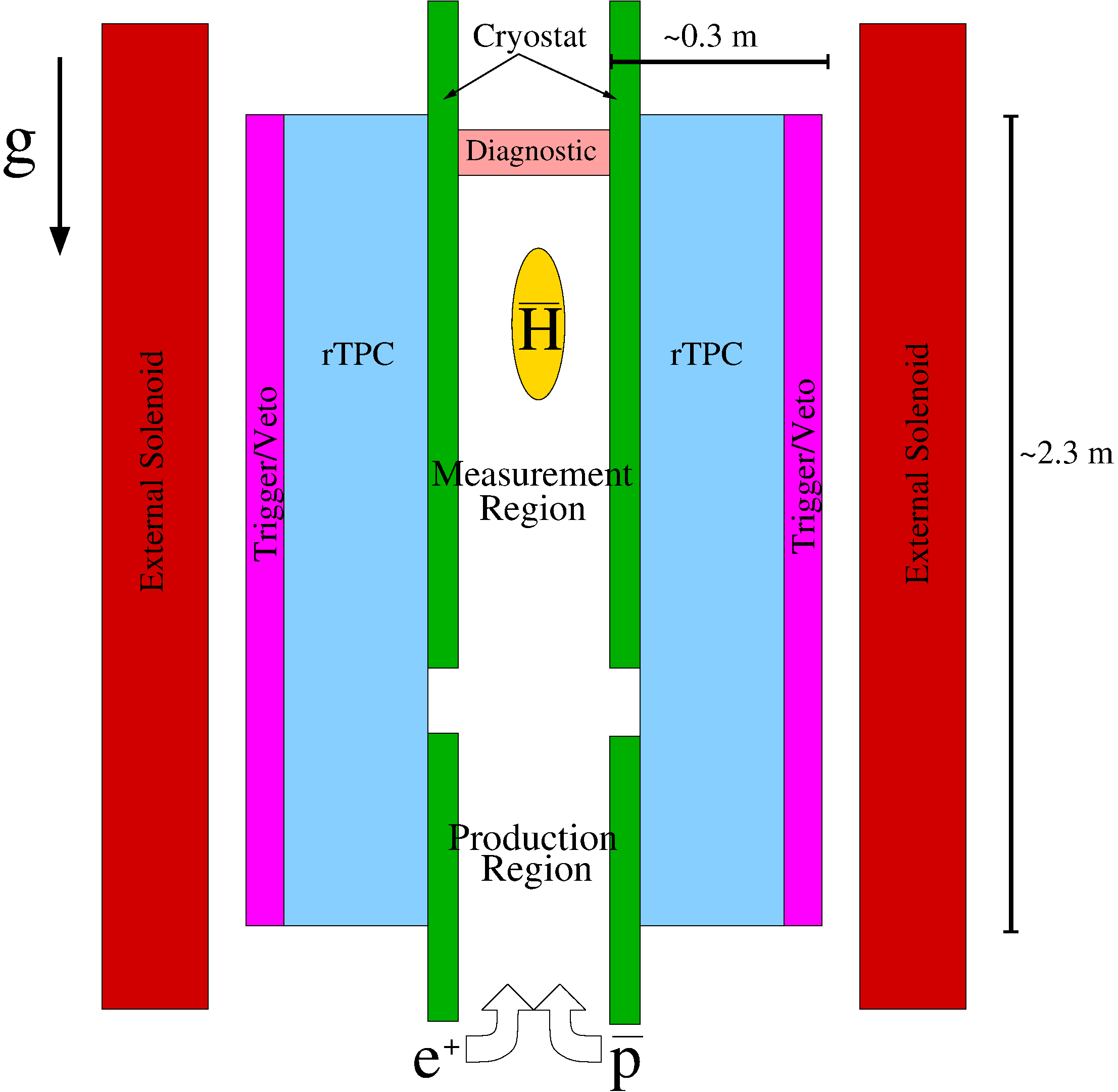}
 \caption{Sketch of ALPHA-g. The volume labelled ``cryostat'' (green) stands for the materials listed in Tab.~\ref{tab:A-g_MC}. The $\h$ constituents, antiprotons $\anti{p}$ and positrons $\pos$, are injected into the apparatus from the bottom. The external solenoid (red) provides the uniform magnetic field to confine charged particles and antiparticles in a Penning trap. The ``trigger/veto'' system (pink) is constituted by an array of scintillating bars that helps to reject background events due to cosmic rays.}
 \label{fig:Ag}
\end{figure}

The ALPHA collaboration, based at CERN-AD, has proposed a novel experiment, called \textit{ALPHA-g}, that aims to measure the gravitational acceleration of $\h$. The ALPHA-g apparatus is \textit{vertical}, i.e., its axis is parallel to the Earth's gravitational field, with the $\anti{p}$ and the $\pos$ injected into the Penning trap from the bottom end. The apparatus is divided into two regions, at different heights, with different functions (see Fig.~\ref{fig:Ag}). The lower part, called the trapping or mixing region, is similar to the ALPHA mixing region \cite{am14aata}, where an electrode stack, together with a solenoid magnet, are used for $\anti{p}$ and $\pos$ manipulation, leading to the formation and trapping of cold $\h$. This is achieved by means of a superconducting octupole magnet, that provides the radial confinement, and a set of coils, called \textit{mirror} coils, that provides the axial confinement. The upper part, called the analysis or measurement region, is where the actual determination of the $\h$ gravitational acceleration takes place in two stages: first its ``sign'' (i.e., ``up'' or ``down'') and later with $1\%$ accuracy. In order to achieve the latter, precise magnetometry and magnets control are particularly important aspects of the ALPHA-g design, since magnetic field gradients, which are used to confine $\h$, can also mimic gravity.

A fundamental requirement to perform gravity measurements with the ALPHA-g apparatus is a \textit{position sensitive particle detector}, which provides information about $\h$ annihilation location, called the \textit{vertex}. The technology of choice is a Time Projection Chamber, or TPC, that is a gas detector capable of imaging the $\h$ annihilation by sampling the annihilation products multiple times (hits). These track samples contain three-dimensional information of the track thus are called \textit{spacepoints}.

\section{ALPHA-\MakeLowercase{g} \MakeLowercase{r}TPC}
The main function of the ALPHA-g \textit{radial TPC} - \textit{rTPC} - is to identify $\h$ annihilation for monitoring the $\h$ production and confinement, as well as to perform the measurement of its gravitational acceleration. This includes discrimination between $\h$ annihilation and cosmic rays, the main source of background in the ALPHA-g experiment.

\begin{figure}[h!]
 \centering
 \includegraphics[scale=0.1]{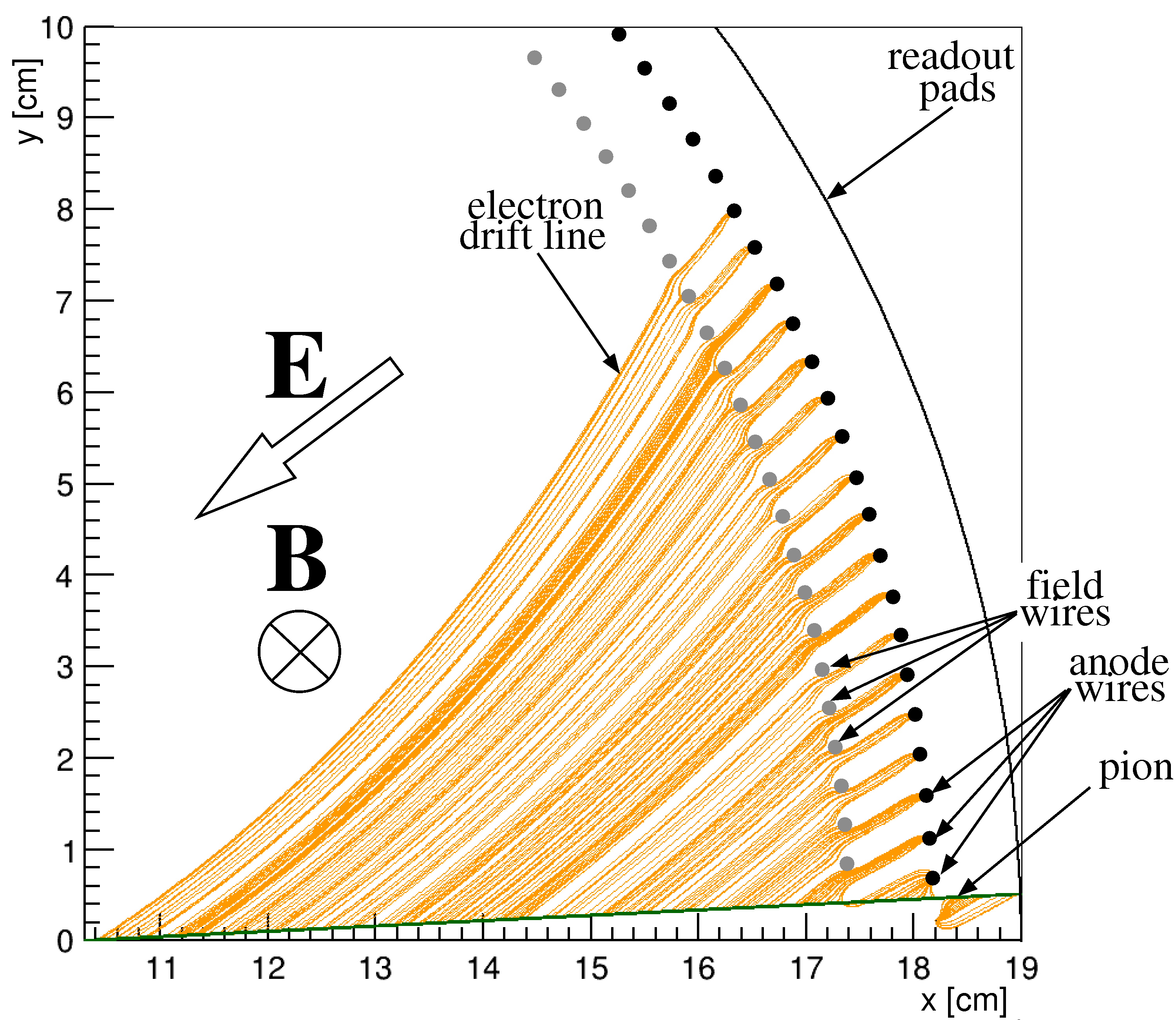}
 \caption{Cross-section view of a wedge of the rTPC. $\pic$ (green line) produces $\ele\mathrm{-ion}$ pairs in the gas. The $\ele$ (orange lines) are drifted towards the anode wires under the influence of a radial electric field and of an axial magnetic field ($1\,$T). The size of the field and anode wires is exaggerated for clarity's sake. The drift lines are obtained with Garfield++ \cite{garfpp}.}
 \label{fig:cell_drift}
\end{figure}

The rTPC consists of a large cylinder ($\sim2.3\,$m long, $\sim40\,$cm outer diameter and \mbox{$\sim20\,$cm} inner diameter) filled with a mixture of $\text{Ar}$ and $\text{CO}_2$ gases at room temperature and atmospheric pressure. The chamber readout is performed at the outermost radius (thus the name ``radial'' TPC), where the TPC is terminated with a multi-wire proportional chamber (MWPC), consisting of anode and field wires, as shown in Fig.~\ref{fig:cell_drift}. The outer wall of the MWPC consists of a printed circuit board (outer cathode) segmented in pads. The electrons released by the ionization radiation in the gas volume drift towards the anode wires, where they are collected and amplified, producing a negative signal on the anodes and inducing signals on the facing pads \cite{sa14grd}.

Two features distinguish the rTPC from other TPCs: the solenoidal magnetic field used to confine the charged antiparticles is orthogonal to the electric field used to drift the ionization in the chamber and the tracks are bound to cross the pad plane. The former feature requires to correct the hits position for a quantity known as the \textit{Lorentz angle}, that is the angle between the $\ele$ drift velocity and the magnetic field.

The rTPC detects the $\anti{p}$ annihilation products, neglecting the 2 $\gamma$ from the $\pos$ annihilation. The main $\anti{p}$ annihilation mode is into several $\pic$ and $\pio$, the latter decaying in $\sim10^{-17}\,$s to two $\gamma$. The $\pic$ trajectories - \textit{tracks} - are reconstructed from hit positions. The $\h$ annihilation vertex is determined by finding the point where the reconstructed tracks pass closest to one another.

Particle tracking in the ALPHA-g environment is challenging: the distance between the annihilation point and the first measurement of the annihilation products is of the order of $10\,$cm and the abundance of material with high density causes a degradation of the tracking performance due to multiple scattering. The main parameter that governs such a process is the radiation length of the material. For the whole ALPHA-g apparatus this is approximately $3\,\mathrm{cm}$. For a $\pic$ meson with momentum magnitude of $250\,$MeV, travelling for $10\,$cm, the RMS of the distribution of scattering angle \cite{ol14rpp} is $5^{\circ}$, causing a displacement with respect to the original path, hence from the true annihilation point, of $7\,$mm. In addition, the substantial amount of scattering material (with high atomic number Z) provides also the ideal environment for the high energy photons, produced in the instantaneous decay of the $\pio$ meson, to create $\ele$-$\,\pos$ pairs. Since these particles do not directly point back to the antihydrogen annihilation point, they add a challenge to the vertex determination.

\section{Simulation}
The Geant4 \cite{ag03gst,al06gda} simulation of the TPC for ALPHA-g was initially used to study the accuracy with which the $\h$ annihilation vertex is reconstructed as a function of the radius of the drift region, i.e., the active volume where the annihilation products can ionize the gas and the ionization products drift to readout pads. Additional simulations were performed to study the behaviour with different anode wires spacing, pad sizes and timing resolution, which determine the hit accuracy. Since it is necessary to establish the TPC performance by developing a reconstruction algorithm, it is required to fix the aforementioned TPC parameters in order to tune the various tracking and vertexing parameters. The chosen configuration corresponds to a drift radius of \mbox{$R_{\text{TPC}}=10\,$cm}, a pad size of $4\,$mm in both the axial direction and anode wires angular spacing of $1.4^{\circ}$, corresponding to $\sim4\,$mm. The following refers specifically to this case.

In the absence of a faithful simulation of the readout, including the electron multiplication at the anode wires, the present Monte Carlo has to take into account the minimum separation in time between two hits, i.e., the minimum time for which two drifting electrons are distinguishable. This characteristic determines the accuracy with which the radial position of the spacepoints is reconstructed. In the following such a readout time resolution is assumed to be $10\,$ns. The main effect of increasing this value is to reduce the number of reconstructed spacepoints along the tracks, so that the overall reconstruction accuracy is slightly decreased.

The inner wall of the TPC, made of polycarbonate, is at a radius of $10\,$cm and it is $3\,$mm thick. The drift region extends over the radius $R_{\text{TPC}}$ and contains a gas mixture of $90\%\,\mathrm{Ar}$ and $10\%\,\mathrm{CO}_2$. The outer wall of the TPC at \mbox{$10.3\,\mathrm{cm} + R_{\text{TPC}}$} is included in the Monte Carlo but is irrelevant for the current purpose.
\begin{table}[h!]
 \centering
 \begin{tabular}{|l|l|l|l|}
  \hline
  Name 			&Radius [cm]	&Material		&Note\\
  \hline
  Electrodes		&2.2275		&Al			&Part of the Penning trap.\\
  UHV chamber		&2.3775		&stainless steel	&\\
  \hline
  Octupole		&2.5025		&Cu-Nb-Ti		&Superconducting magnet for $\h$ radial confinement.\\
  Mirror coils		&3.9025		&Cu-Nb-Ti		&Superconducting magnets for $\h$ axial confinement.\\
  \hline
  OVC inner wall	&5.5		&stainless steal	&\\
  \hline
  Liquid He space	&\multicolumn{2}{l|}{between UHV and OVC}&Cooling for superconducting magnets.\\
  \hline
  Heat shield				&6.2		&Cu	&\\
  OVC outer wall	&7.6		&stainless steel	&\\
  \hline
 \end{tabular}
 \caption[Material budget in ALPHA-g MC.]{Material budget in ALPHA-g Monte Carlo. ``UHV'' stands for Ultra-High-Vacuum, while ``OVC'' for Outer-Vacuum Chamber. The radius column refers to the inner one of the volume.}
 \label{tab:A-g_MC}
\end{table}

The present simulation includes a $1\,$T solenoidal field, along the vertical axis, $z$, while no attempts have been made to introduce the detailed magnetic field of ALPHA-g, since it has not been clearly defined yet.

The $\h$ annihilation is simulated as the process $\h + \mathrm{p} + \ele \rightarrow n\pic + m\pio + 2\gamma$, where $n$ and $m$ are the ``pions multiplicities'' as measured in \cite{bende}.

The Geant4 simulation begins with the generation of $\h$ annihilation uniformly on the trap wall, i.e., on a circumference of radius \mbox{$R_w = 22.275\,$mm}. Axially, i.e., along the $z$ axis, the annihilations are distributed uniformly over a length of $2\,$cm centred around \mbox{$z=50\,$cm}. The $\h$ annihilation is assumed to occur at rest so that the total energy available for the annihilation products is twice the mass of the proton $\approx1.9\,$GeV.

The gas ionization due to the passage of $\pic$ in the drift chamber is simulated with the \textit{PhotoAbsorption Ionization}, or PAI, model that ``describes the ionization energy loss of a relativistic charged particle in matter'' \cite{ap00iielvtagsp}.

\begin{figure}[h!]
 \centering
 \includegraphics[scale=0.18]{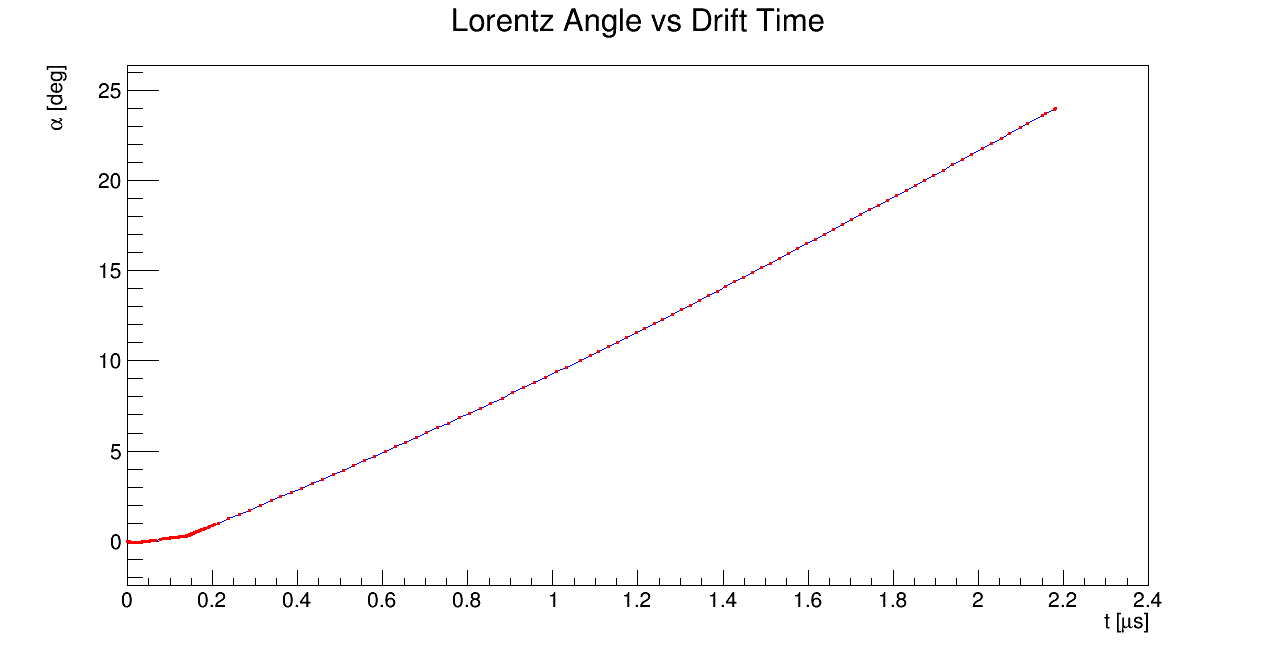}%
 \includegraphics[scale=0.18]{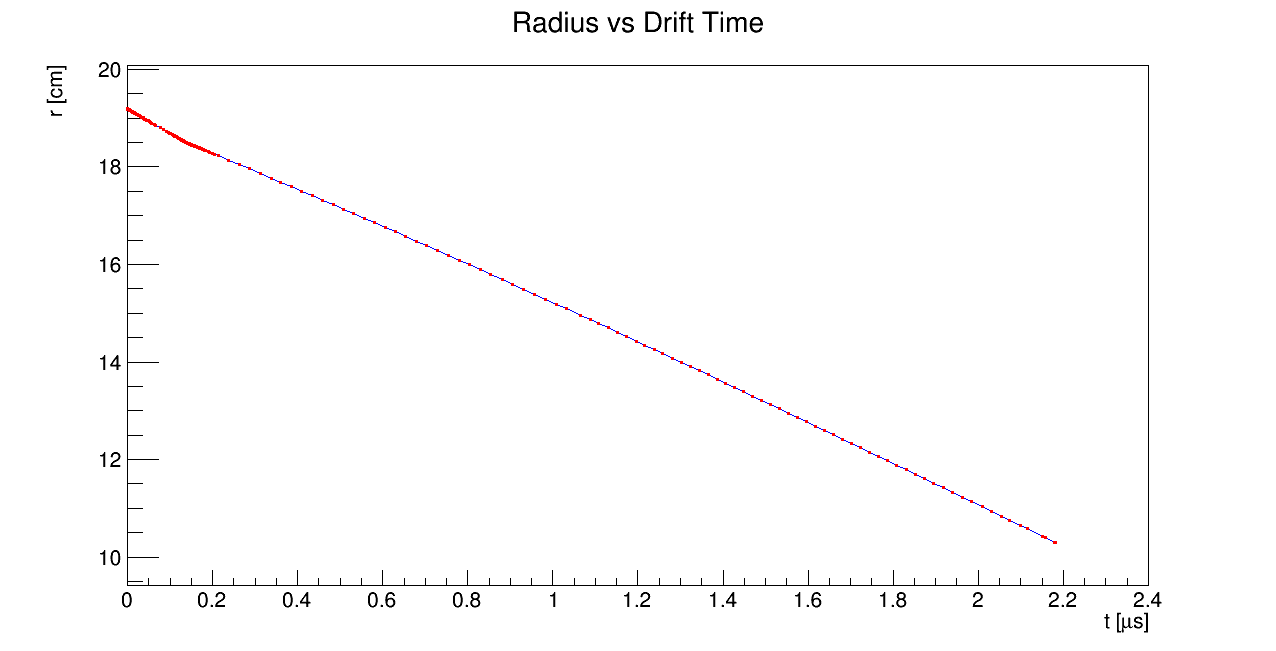}
 \caption[Lorentz angle $\alpha$, drift time and hit radius.]{Lorentz angle $\alpha$ and drift time $t_d$ experienced by the $\ele$ as a function of the radius where they are created. The data used to produce these plots are obtained with Garfield \cite{garf}.}
 \label{fig:e-drift}
\end{figure}

The position and the time of the primary ionization is recorded and \textit{digitized} from Monte Carlo information and \textit{ad-hoc} Garfield calculation \cite{garf}, shown in Fig.\ref{fig:e-drift}.

The digitization routine saves the triplets to a ROOT \cite{ROOT} file ready for the reconstruction.

\section{Vertex Reconstruction}
With the Monte Carlo data at hand it is time to turn to the reconstruction routines and evaluate the performance of the algorithm. The position of a spacepoint in the rTPC is given by the three cylindrical coordinates: $r$ from the $\ele$ drift time $t_d$, extracted from the anode wire hit, $\varphi$ from the anode wire number, and corrected by the \textit{Lorentz angle} $\alpha$, and $z$ from the pad number.
 
A set of hits belonging to a single track form a specific pattern that must be identified in order to use such a set in an algorithm that calculates the track parameters. A so-called \textit{pattern recognition} algorithm takes as input the whole set of spacepoints in each annhilation event and produces sets of points belonging to different tracks. A track is identified by searching nearby spacepoints within a sphere of fixed radius, starting from points at the largest possible radius.

With the presence of a uniform magnetic field, the trajectory of the $\pic$ can be modelled by an helix, whose \textit{canonical} form can be found in \cite{av92aft}. A Least Squares fitting of the helix to the hits is performed by minimizing the $\chi^2$ figure of merit. Since not all the reconstructed tracks are due to $\pic$ and, most importantly, not all of them point back to the $\h$ annihilation vertex (e.g., low-energy electrons from photon conversion), it is convenient to remove such helices by placing a cut-off on the normalized $\chi^2$ and on other helix parameters, such as the curvature. Helices satisfying the default cuts are dubbed \textit{good helices}.

Since the charged particles travel through several centimetres of dense material before reaching the detector, a more accurate account of the errors on the reconstructed tracks is desirable. Multiple scattering causes the particle to scatter in the two planes perpendicular and parallel to its path without losing energy. The distribution of each angle is Gaussian with standard deviation \cite{av92aft} $\sigma_{\text{MS}}=\tfrac{0.0141}{p \beta}\sqrt{\tfrac{L}{X_R}}$, where $L$ is the path length, \mbox{$X_R\approx 3.21\,$cm} is the radiation length given, $\beta$ is the velocity and $p$ is the momentum in GeV/$c$. The errors from the best-fit and the multiple scattering are added in quadrature $\sigma^2=\sigma_{\text{fit}}^2 + \sigma_{\text{MS}}^2$.

The vertex reconstruction is achieved through a three step procedure. The first step, or \textit{seeding}, consists in finding the minimum distance between each pair of good helices. The \textit{seed vertex} is taken as the midpoint of the segment joining the pair of helices that pass closest to each other. The calculation of the midpoint takes into account the errors associated with the minimum distance points. In the second step, a \textit{recalculation} of the vertex position is performed by minimizing the distance of the best pair of helices to a common point, called the \textit{recalculated vertex}. Initial values for such a minimization are taken from the previous step. The last step is only possible if the number of good helices is greater than two, aiming to \textit{improve} the vertex resolution by minimizing the distance of the best pair of helices and additional helices to a common point. Every time a new helix is added to the set participating in the minimization to the common point, the newly calculated $\chi^2$ is required to be less than a cut-off value. Additional details on the reconstruction methodology can be found in \cite{thesis}. 
\begin{figure}[h!]
 \centering
 \includegraphics[scale=0.27]{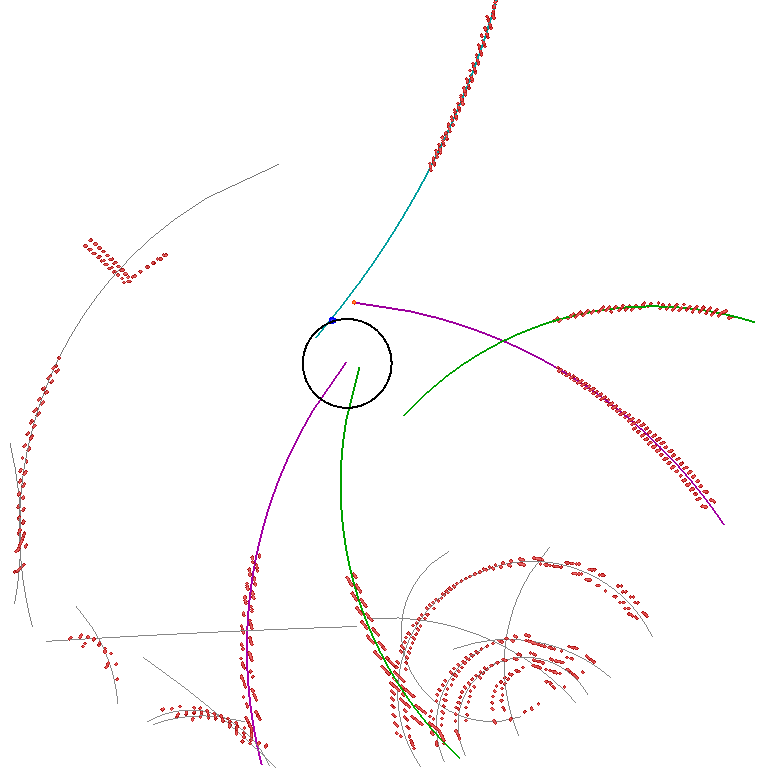}%
 \includegraphics[scale=0.32]{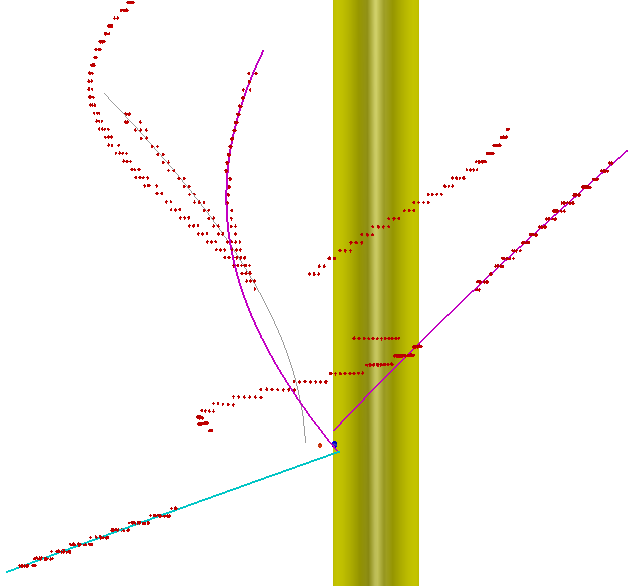}
 \caption[Reconstructed event.]{Reconstructed events. The red dots are the reconstructed spacepoints. The purple lines indicate helices participating in the reconstruction of the seed vertex. The azure lines are helices added to the vertexing in the improvement step. Green lines are ``good helices'' but not used for vertexing. Grey lines are helices that failed the cuts selection.  The Monte Carlo vertex is the blue dot, while the reconstructed one is orange. Left: front view, where the electrode stack is represented by the black ring. Right: side view, where the electrode stack is represented by the gold tube. The electrode stack has a radius of $22.275\,$mm.}
 \label{fig:tracks02}
\end{figure}

\section{Preliminary Results}
The \textit{vertexing efficiency}, which is the number of events where the vertex is successfully found over the total number of generated events, is $94\%$ and depends on the optimization of the cuts on the $\chi^2$ and curvature given by the best-fit to the helices, since the vertex position can be obtained as long as there are at least two reconstructed helices.

The position of the reconstructed vertex $(r_v, \phi_v, z_v)$ is compared to the actual point of origin of the $\pi^{\pm}$, the so called Monte Carlo vertex \mbox{$(r_{\text{MC}}=22.275\,\mathrm{mm}, \phi_{\text{MC}}, z_{\text{MC}})$}. The distributions of \mbox{$r_{\text{MC}} - r_v$}, \mbox{$\phi_{\text{MC}} - \phi_v$} and \mbox{$z_{\text{MC}} - z_v$} are fitted to a Gaussian in a restricted interval around the centroid and the estimated $\sigma$ is taken as detector resolution, as shown in Tab.~\ref{tab:res}.
\begin{table}[h!]
 \centering
 \begin{tabular}{|c|c|}
  \hline
  \multicolumn{2}{|c|}{\textbf{Vertex Resolution}}\\
  \hline
  $r$ 		&$7\,\mathrm{mm}$\\
  $\phi$ 	&$15^{\circ}$\\
  $z$ 		&$4\,\mathrm{mm}$\\
  \hline
 \end{tabular}
 \caption{Vertex resolution in cylindrical coordinates: the reported value is the standard deviation of the best-fit Gaussian in a restricted interval around the centroid of the underlying distribution.}
 \label{tab:res}
\end{table}

\section{Conclusions}
The first part of the present section describes a realistic (but not complete) simulation of an apparatus, called ALPHA-g, designed to measure the gravitational interaction of $\h$. The simulation is focused on the Time Projection Chamber intended to determine the location of the $\h$ annihilation upon release from the trap or interaction with radiation.

The second part of the present document describes the full reconstruction software for the ALPHA-g TPC. From the digitized signals on pads and anodes, produced by the primary ionization, the $\h$ annihilation position is inferred by tracking the annihilation products back to the common origin. 

The tracks are determined by finding where the charged particles produced the primary ionization, called spacepoints, and by identifying the common pattern underlying the spacepoints envelope of a track. The set of spacepoints belonging to a single charged particle are best-fitted with a helix, in order to take into account the solenoidal magnetic field, using the least-squares method. The helices are selected based on the best-fit result and the ones passing the selection criteria are used to calculate the annihilation position, namely the vertex.
%

\end{document}